\documentclass[slac_one]{revtex4}
\usepackage{graphicx}
\usepackage{fancyhdr}
\usepackage{epsfig}
\usepackage{epstopdf}
\begin{document}
\title{\textit{W/Z}+jets as signal and background in ATLAS}
\author{E. Dobson on behalf of the ATLAS collaboration, e.dobson1@physics.ox.ac.uk}
\affiliation{Particle Physics Department, Denys Wilkinson Building, Keble Road, Oxford}

\begin{abstract}
Events containing \textit{W} or \textit{Z} bosons with accompanying jets are important channels to test the Standard Model. These proceedings discuss \textit{W/Z}+Jets production cross section measurements and their use to understand the detectors and their performances after the LHC start-up. At the same time, these processes are a major source of background to new physics searches, such as SUSY. Methods, mostly data-driven, to identify these events as signals and backgrounds to searches will be presented.
\end{abstract}

\maketitle
\thispagestyle{fancy}
\section{Introduction}
In this note, we examine the triggering, reconstruction and analysis of events containing a \textit{Z} boson plus jets in ATLAS \cite{AtlasDetector}. We concentrate on the \textit{Z} boson's decay into electrons and muons. Comparisons of the reconstructed/corrected quantities, when possible, are made to truth-level hadron information, and also to parton-level information with parton-to-hadron corrections applied. Our primary end-result are hadron-level cross sections, similar to what we expect to have with the real ATLAS data. We present expectations/yields/systematics scaled to 1fb$^{-1}$, an integrated luminosity that should be accumulated within the first two years of running, but we will comment on difficulties expected and analysis strategies to be adopted during the early running. We study the comparison of theory and measurement expectations for quantities suited to compare with a fixed-order NLO calculation: in this note the inclusive cross-section for \textit{Z}$\rightarrow$\textit{ll} with at least 1 jet, 2 jets and 3 jets. Discussion of expected systematic errors is included.\\
\indent \textit{W/Z}+jets are not only interesting in their own right but also form a significant predicted background to many new physics channels and it will be impossible to discover, for example, a Higgs Boson or SUSY particles unless the backgrounds are well understood. Due to incomplete knowledge regarding many input variables (underlying event, parton showering, cross sections, pdfs, calibration etc) it is wise to try and measure these backgrounds from data driven methods. To detail all of these would be beyond the scope of this note and thus two particular analyses (Higgs and SUSY) are elaborated in more detail here.

\section{\textit{W/Z} + JETS AS SIGNAL}
\begin{figure*}[h]
\centering
\includegraphics[scale=0.4]{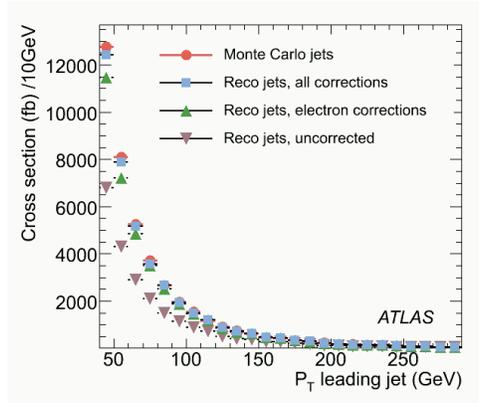}
\caption{Unfolding \textit{Z}$\rightarrow$\textit{ee} jet distribution from detector to truth level.} 
\label{SignalPlots1}
\end{figure*}

\textbf{Experimental studies}\\
\indent A cross section measurement requires the reconstruction and trigger efficiencies to be known. The primary approach to measure these for \textit{W} or \textit{Z} lepton decay analysis in ATLAS will most likely be the data-driven `Tag and Probe' method \cite{CSC} (electron trigger chapter) in \textit{Z}$\rightarrow$\textit{ll} events. The presence of jets in the \textit{Z}+jet events is expected to affect the measured efficiencies, and this will have to be taken into account. \\ 
\indent For comparison with theory, reconstructed data have to be unfolded from the detector level to the hadron level, correcting for the reconstruction and trigger efficiencies, as well as resolution and non linearities in electron and jet reconstruction. The impact of these corrections may be seen in figure \ref{SignalPlots1}(right), where the dominant correction stems from the electron reconstruction. For a cross section measurement, the main detector uncertainty originates from the jet energy scale (JES), expected to start from an uncertainty at the level of 10\% and converging towards 1\%. A JES uncertainty of 10\% would be the dominant error on the cross section (15-30\%). This is at the same order as the typical differences expected between LO and NLO predictions, or between predictions from PYTHIA \cite{pythia}, ALPGEN \cite{alpgen} and MCFM \cite{MCFM}. \\\\

\begin{figure*}[ht]
\centering
\includegraphics[scale=0.55]{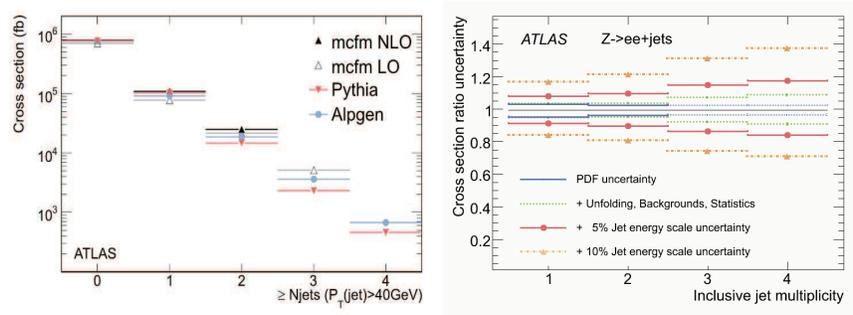}
\caption{(left) Comparisons of MCFM and generator predictions of jet multiplicity in \textit{Z}$\rightarrow \mu\mu$. (right) Dominant uncertainties on the cross section measurement.} \label{SignalPlots2}
\end{figure*}

\textbf{Theoretical studies}\\
\indent Cross section comparisons have been made between MCFM predictions and PYTHIA/ALPGEN (interfaced with HERWIG), where MCFM predictions have been corrected to the hadron level. The differences may be seen in figure \ref{SignalPlots2}(left). In general, PYTHIA is seen to predict a softer P$_T$ spectrum than ALPGEN, although both agree reasonably well with MCFM predictions. The main theoretical uncertainty on a boson+jets cross section measurement has been seen to be the PDF uncertainty, which has been shown as a comparison to the JES uncertainty in figure \ref{SignalPlots2}(right).

\section{\textit{W/Z} + JETS AS BACKGROUND}
\subsection{SUSY}
\begin{figure*}[ht]
\centering
\includegraphics[scale=0.6]{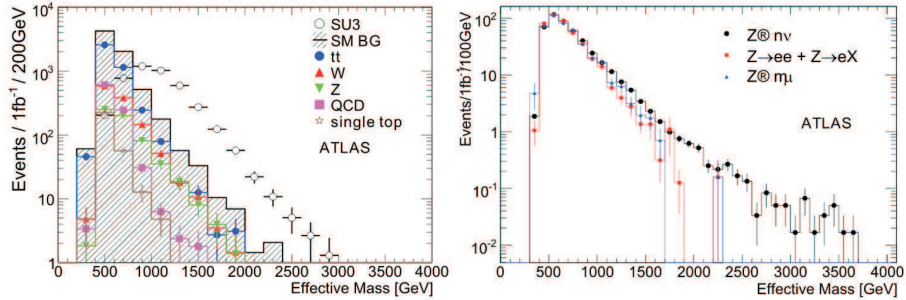}
\caption{(left) Signal and background in the SUSY no lepton channel. (right) Estimated background for the SUSY no lepton channel after using \textit{Z}$\rightarrow$\textit{ll} to estimate the \textit{Z}$\rightarrow \nu\nu$ contribution.}  \label{SUSY}
\end{figure*}

\textit{W} or \textit{Z}+jets are a major background \cite{CSC} (SUSY chapter) to several SUSY search channels, many of which require leptons, jets and missing transverse energy (MET) in their event selection. It is wise to estimate these backgrounds using data driven methods. The accepted strategy in ATLAS SUSY analyses is to derive such a prediction from a control `SUSY free' region and extrapolate to the signal region. \textit{W}+jets production is a major background in the one lepton channel, whilst \textit{Z}+jets must be considered for the no lepton channel. \\
\indent The cuts for this search mode are on $\geq$4 jets and large MET: \textit{Z}$\rightarrow \nu\nu$+jets is a major background, as may be seen in figure \ref{SUSY}(left). It is estimated by scaling (as according to equation \ref{SUSYScaleEq} where the \textit{c} factors account for kinematical and fiducial differences) a control sample obtained by using standard \textit{Z}$\rightarrow$visible lepton event selection. Given the similar event kinematics to the neutrino sample, this may be used to estimate the MET and effective mass distributions, as may be seen in figure \ref{SUSY}(right). The resultant dominant systematics are the variation of the renormalisation scale in ALPGEN and the soft part of the MET which isn't taken into account when replacing neutrinos by leptons. 

\begin{equation}\label{SUSYScaleEq}
N_{Z\rightarrow\nu\nu}(ME_t)=N_{Z\rightarrow ll}(P_t(l^+l^-)\times c_{kin}(P_t(Z))\times c_{fidu}(P_t(Z))\times\frac{BR(Z\rightarrow\nu\nu)}{BR(Z\rightarrow l^+l^-)}
\end{equation}

\subsection{Higgs}
\begin{figure*}[ht]
\centering
\includegraphics[scale=0.5]{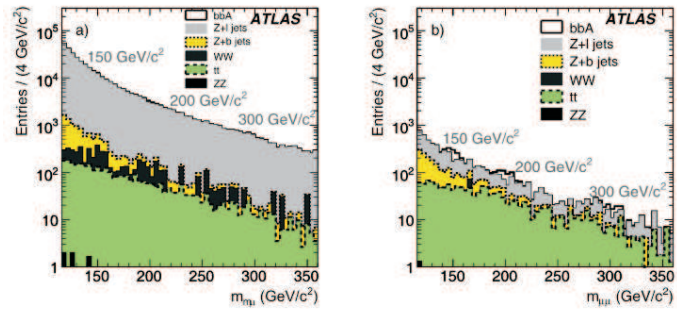}
\caption{Signal and background for the MSSM Higgs \textit{A/H/h}$\rightarrow \mu^+\mu^-$ channel: (left) without \textit{b} tagging (right) with \textit{b} tagging.} \label{Higgs}
\end{figure*}

\textit{W/Z}+jets are relevant for many Higgs channels, and the MSSM Higgs \textit{A/H/h}$\rightarrow \mu^+\mu^-$ channel is given as an example. The selection for this process requires two isolated muons and small MET and thus \textit{Z} decaying to muons in association with jets is a background. The \textit{Z}+light jet contribution to the background is cut down using \textit{b} tagging, as may be seen in figure \ref{Higgs}. The \textit{b} tagging weights (in combination with MET and \textit{b} jet multiplicity) may be used as a discriminating variable to separate signal and background. 

\end{document}